# Direct Observation of Layer-Dependent Electronic Structure in Phosphorene

Likai Li[1,5†], Jonghwan Kim[2†], Chenhao Jin[2†], Guojun Ye[3,4,5], Diana Y. Qiu[2], Felipe H. da Jornada[2], Zhiwen Shi[2], Long Chen[7], Zuocheng Zhang[1,5], Fangyuan Yang[1,5], Kenji Watanabe[6], Takashi Taniguchi[6], Wencai Ren[7], Steven G. Louie[2,8*], Xianhui Chen[3,4,5*], Yuanbo Zhang[1,5*] and Feng Wang[2,8,9*]

[1]State Key Laboratory of Surface Physics and Department of Physics, Fudan University, Shanghai 200433, China.

[2]Department of Physics, University of California at Berkeley, Berkeley, California 94720, USA.

[3]Hefei National Laboratory for Physical Science at Microscale and Department of Physics, University of Science and Technology of China, Hefei, Anhui 230026, China.

[4]Key Laboratory of Strongly Coupled Quantum Matter Physics, University of Science and Technology of China, Hefei, Anhui 230026, China.

[5]Collaborative Innovation Center of Advanced Microstructures, Nanjing 210093, China.

[6]Advanced Materials Laboratory, National Institute for Materials Science, 1-1 Namiki, Tsukuba, 305-0044, Japan

[7]Shenyang National Laboratory for Materials Science, Institute of Metal Research, Chinese Academy of Sciences, Shenyang 110016, China

[8]Materials Sciences Division, Lawrence Berkeley National Laboratory, Berkeley, California 94720, USA.

[9]Kavli Energy NanoSciences Institute at the University of California, Berkeley and the Lawrence Berkeley National Laboratory, Berkeley, California, 94720, USA.

† These authors contributed equally to this work.

\* Email: fengwang76@berkeley.edu; zhyb@fudan.edu.cn; chenxh@ustc.edu.cn; sglouie@berkeley.edu

**Abstract:**

**Phosphorene, a single atomic layer of black phosphorus, has recently emerged as a new two-dimensional (2D) material that holds promise for electronic and photonic technology. Here we experimentally demonstrate that the electronic structure of few-layer phosphorene varies significantly with the number of layers, in good agreement with theoretical predictions. The interband optical transitions cover a wide, technologically important spectrum range from visible to mid-infrared. In addition, we observe strong photoluminescence in few-layer phosphorene at energies that match well with the absorption edge, indicating they are direct bandgap semiconductors. The strongly layer-dependent electronic structure of phosphorene, in combination with its high electrical mobility, gives it distinct advantages over other two-dimensional materials in electronic and opto-electronic applications.**

Atomically thin 2D crystals have emerged as a new class of materials with unique material properties that are potentially important for electronic and photonic technologies[1–10]. Various 2D crystals have been uncovered, ranging from metallic (and superconducting) $NbSe_2$ and semimetallic graphene to semiconducting $MoS_2$ and insulating hexagonal boron nitride (hBN). The energy bandgap, a defining characteristic of an electronic material, varies correspondingly from 0 (in metals and graphene) to 5.8 eV (in hBN) in these 2D crystals. Despite the rich variety currently available, 2D materials with a bandgap in the range from 0.3 eV to 1.5 eV are notably missing[11]. Such a bandgap corresponds to a spectral range from mid-infrared to near-infrared that is important for optoelectronic technologies such as telecommunication and solar energy harvesting. It is therefore desirable to have 2D materials with a bandgap that falls in this range, and in particular, matches that of the technologically important silicon (bandgap = 1.1 eV) and III-V semiconductors like InGaAs, without compromising sample mobility[12].

Monolayer and few-layer phosphorene are predicted to bridge the much needed bandgap range from 0.3 to 2 eV (Refs. 13–17). Inside monolayer phosphorene, each phosphorus atom is covalently bonded with three adjacent phosphorus atoms to form a puckered honeycomb structure[18]. The three near $sp^3$ bonds together with the lone-pair orbital take up all five valence electrons of phosphorus, so monolayer phosphorene is a semiconductor with a predicted direct optical bandgap of ~ 1.5 eV at the Γ point of the Brillouin zone. The bandgap in few-layer phosphorene can be strongly modified by interlayer interactions, which leads to a bandgap that decreases with phosphorene film thickness, eventually reaching 0.3 eV in the bulk limit. Experimental observations of layer-dependent band structure in phosphorene, on the other hand, have been rather limited. Previously, photoluminescence (PL) spectroscopy has been used to probe the bandgap of monolayer and few-layer phosphorene[8,19–22]. Such studies, however, have their limitations in that PL can be dominated by defect and impurity states rather than the bandgap emission and that few-layer phosphorene is highly susceptible to degradation. Indeed, different studies have reported widely different bandgap values from 1.32 to 1.75 eV for monolayer phosphorene[8,20–22].

Here we report the first study of optical absorption in high quality few-layer phosphorene on a sapphire substrate with an hBN capping layer of ~15 nm thickness. The optical absorption is insensitive to defects and impurities levels, which complicate PL measurements. The absorption spectroscopy therefore provides a reliable determination of the evolution of the optical bandgap in monolayer and few-layer phosphorene. We found the bandgap of monolayer, bilayer, and trilayer phosphorene (sandwiched between sapphire and hBN) to be 1.73, 1.15, and 0.83 eV, respectively, and bulk black phosphorus has a bandgap of 0.35 eV. Comparisons with PL of the same samples show that PL peaks are close to the absorption bandgap, confirming the direct nature of the bandgaps. Interestingly, the bandgap of bilayer and trilayer phosphorene matches well with that of silicon (1.1 eV) and telecom photon energy (0.8 eV), respectively. In addition, optical absorption above the bandgap reveals extra resonances in few-layer phosphorene corresponding to transitions between higher sub-bands arising from quantum confinement along the thickness directions. The systematic evolution of both the optical bandgap and absorption peaks arising from higher sub-band transitions matches well with *ab initio* GW Bethe Salpeter equation (GW-BSE) calculations and can be captured by a simple, phenomenological tight-binding model. The layer-tunable electronic structure of phosphorene can be further varied through electrostatic field and mechanical strain[23,24], opening up exciting possibilities for both electronic devices and optoelectronic applications from the infrared to visible spectral range.

Monolayer and few-layer phosphorene samples are prepared by micromechanical exfoliation of bulk crystals. To avoid sample degradation in air, all samples are fabricated in an inert gas glove box with oxygen and moisture levels lower than 1 ppm. We first exfoliate phosphorene flakes onto silicon substrates with a 300 nm oxide layer and identify few layer samples using optical microscopy (Fig. 1b and c). Sample thickness was first estimated from the optical contrast under microscope, and also from the apparent height determined by atomic force microscopy (AFM; see supplementary information). We found that with each additional atomic layer, the optical contrast in the red channel of a color CCD camera increases by ~ 7 percent in few-layer phosphorene on $SiO_2$/Si substrate (Fig. 1d). Accurate thickness determination for isolated phosphorene flakes is difficult using AFM due to the different tip-surface interaction on phosphorene and $SiO_2$/Si wafer. However, the AFM height difference between two adjacent flakes of different thickness can yield useful information, since an average height increase of 0.5 nm is expected for one additional atomic layer (supplementary information). Such estimation of phosphorene layer number is further verified through optical spectroscopy results of the flakes, as we will describe later. To facilitate optical spectroscopy and avoid degradation, we transferred the phosphorene flakes to a sapphire substrate with a top hBN encapsulation using a dry transfer technique in the glove box[25]. The thickness of the hBN flake is around 15 nm. The fabricated samples were kept in high vacuum in an optical cryostat during spectroscopic studies.

We probed optical absorption of monolayer, bilayer, trilayer, tetralayer, and pentalayer phosphorene through reflectance measurements at 77 K. The reflection spectrum $\Delta R/R$ is directly related to the complex dielectric function of phosphorene (supplementary information). The imaginary part of the dielectric function is proportional to optical absorption and features prominent absorption peaks, while the real part of the dielectric function can lead to relatively broad backgrounds. In this study, we will focus on the optical resonances, where the position of optical absorption peaks can be reliably identified as resonances in the reflection spectra. A combination of supercontinuum laser and tungsten lamp were used as the light source to cover the energy range from near-infrared to visible. The incident light was focused onto the phosphorene flakes in a microscopy setup, and the reflected light was collected and analyzed in a spectrometer equipped with both silicon and InGaAs arrayed detectors. The polarization of incident light was controlled using a broadband calcite polarizer. For comparison, we also measured the absorption spectrum of a thick phosphorene flake (~ 100 nm) using Fourier Transform Infrared Spectroscopy (FTIR) at room temperature, which yields the electronic structure of black phosphorus in the bulk limit.

Figure 2 shows the evolution of polarization-resolved reflection spectra from monolayer to pentalayer phosphorene in the 0.75-2.5 eV spectral range. For monolayer phosphorene, a prominent absorption peak is present at 1.73 eV for $x$-polarized incident light (Fig. 2a, black curve). We attribute the absorption peak at 1.73 eV to the lowest energy exciton arising from transitions at the direct bandgap of monolayer phosphorene because no optical absorption is observed for photon energies below the resonance. This assignment is confirmed by our *ab initio* calculations. In few-layer phosphorene, the optical bandgap decreases continuously with increased layer number: 1.15 eV for bilayer, 0.83 eV for trilayer, and below 0.75 eV (the lower detection limit of our spectral range) for tetralayer and pentalayer (Fig. 2b-e). In bulk black phosphorus, the bandgap reaches 0.35 eV (Fig. 2f). These absorption resonances provide a reliable and quantitative experimental determination of the layer-dependent lowest exciton transition energy or optical bandgap in phosphorene.

All the observed optical absorption spectra exhibit strong anisotropy under *x*- and *y*-polarized illumination (Fig. 2a-f, black vs. red curves). Such strong anisotropy arises from the unique crystal structure of phosphorene, where phosphorus atoms form a puckered honeycomb lattice with completely different behavior along the *x* (armchair) and *y* (zigzag) directions (see Fig. 1a). In particular, phosphorene layers have a mirror symmetry with respect to the *x-z* plane. Electron wavefunctions at the conduction band minimum (CBM) and valence band maximum (VBM) of phosphorene, both at Γ, are composed of *s*, $p_x$ and $p_z$ orbitals with even symmetry with respect to the *x-z* mirror plane. Consequently, optical absorption of *y*-polarized light, which has an odd symmetry with respect to the mirror plane, is strictly forbidden at the bandgap energy due to symmetry requirement. The broad background in reflection spectra is mainly due to contributions from the imaginary part of the optical conductivity, which shows a weaker polarization dependence.

Remarkably, prominent absorption peaks emerge above the optical bandgap in the absorption spectra of few-layer phosphorene. All the above-bandgap optical resonances also exhibit marked polarization anisotropy similar to that of the lowest energy peak, with the peak disappearing when light is polarized along *y*-axis. The high energy resonances also exhibit a systematic layer dependence: the energy of the first above-bandgap resonance (labeled as II) is at 2.44 eV in bilayer, and then decreases to 1.93 eV, 1.52 eV, and 1.16 eV in trilayer, tetralayer, and pentalayer, respectively. The second above-bandgap transition (labeled as III) has a resonance energy of 2.31 eV and 1.94 eV in tetralayer and pentalayer, respectively. These above-bandgap resonances originate from strong layer-layer interactions in phosphorene[17]. In few-layer phosphorene, the quasiparticle states form quantum-well-like states quantized along the thickness direction. For N-layer phosphorene, there will be N quantized states that give rise to N 2D sub-bands from each of the conduction and valence bands of the bulk. In monolayer, bilayer, and trilayer phosphorus, our experimental data and ab initio results show that all the absorption peaks arise from transitions between distinct individual pair of electron and hole subbands, but this picture breaks down in tetralayer phosphorus, where transitions between different higher energy subbands begin to mix considerably. A one-dimensional tight-binding model of the inter-layer interactions successfully describes the bandgap transition energies and higher-energy resonances, which we shall discuss later.

In addition to the optical absorption, we examined PL of the monolayer and few-layer phosphorene at 77 K with unpolarized photoexcitation at 2.33 eV. We observed strong PL close to the absorption bandgap in monolayer, bilayer, and trilayer phosphorene protected with an hBN flake (Fig. 3a-c). It indicates that phosphorene of different layer thicknesses always has a direct bandgap, consistent with theoretical predictions. (The optical bandgap in phosphorene with 4 or more layers is outside the spectral range our detectors.) The PL in monolayer, bilayer, and trilayer show strong polarization dependence similar to that of the absorption spectrum. The PL intensity exhibits a $cos^2\theta$ pattern (Fig. 3d-f) as the light polarization angle $\theta$ varies, consistent with the fact that optical transitions polarized along the *y*-axis are forbidden by symmetry. We note that in all samples, the PL peaks are relatively broad and sometimes contain multiple resonant features, likely due to defect and impurity states that exist even for hBN encapsulated phosphorene samples[26]. The peaks are typically red shifted from the absorption edges by 20 to 40 meV. The red shift in PL could arise from phonon sidebands, photo-induced structural relaxation, and/or defect and impurity states close to the bandgap.

The systematic evolution of both the bandgap transitions and higher-energy resonances in few-layer phosphorene can be described phenomenologically by a one-dimensional tight-binding model for the inter-layer interactions. For simplicity, we will neglect exciton effects (see discussion below) and consider only coupling between single-particle electronic states at the Γ point in the 2D Brillouin zone. The model is therefore described by only three parameters: the bandgap of an isolated monolayer $E_{g0}$ and the nearest-neighbor coupling between adjacent layers for the conduction bands ($\gamma^c$) and valence bands ($\gamma^v$). In an N-layer phosphorene, there are N quantum-well like subbands within both the conduction and valence band complexes, with the Γ-point energy of each subband described by (supplementary information)

$$E_{N,n}^{(s)} = \pm\frac{E_{g0}}{2} - 2\gamma^{(s)} \cos\left(\frac{n}{N+1}\pi\right) \quad (1)$$

where $(s) = c, v$ stands for conduction and valence band complex, and $n = 1\ldots N$ is the sub-band index. Optical transitions are allowed only between conduction and valence bands of the same sub-band indices $n$ for in-plane light polarization (supplementary information), giving rise to optical resonances at $E_{N,n}^R = E_{g0} - 2(\gamma^c - \gamma^v) \cos\left(\frac{n}{N+1}\pi\right)$. Using only two fitting parameters $E_{g0} = 1.8\ eV$ and $\gamma^c - \gamma^v = 0.73\ eV$, the simple tight binding model provides an empirical description of the systematic evolution of the full series of the optical resonances in monolayer and few-layer phosphorene observed in our experiment (Fig. 4a). Our data demonstrate unambiguously the strong layer-layer interactions on the electronic structure of phosphorene. The multiple layer-dependent resonances also provide a reliable way to determine the layer number in phosphorene flakes, and confirm our preliminary assignment based on optical contrast and AFM measurements.

Results from the simple tight binding model are supported by more rigorous *ab initio* GW plus Bethe-Salpeter equation (GW-BSE) calculations, which explicitly include screening[27,28] from the substrate and hBN capping layer as well as self energy[29] and electron-hole interaction effects[30] (see supplementary information for calculation details). Specifically, we find that additional screening due to hBN encapsulation has a much larger effect in phosphorene than in other 2D semiconductors, such as $MoSe_2$[27,28], resulting in a small exciton binding energy of only 80 meV in monolayer phosphorene and even smaller binding energies in few-layer phosphorus[31]. Consequently, the optical resonances are quite close to the interband transition energies in encapsulated phosphorene, justifying our neglecting excitonic effects in the tight binding model. Figure 4b shows the calculated GW quasiparticle bandstructure of encapsulated monolayer, bilayer, trilayer, and tetralayer phosphorene. The calculated optical bandgap energies are 1.56, 0.93, 0.67, and 0.53 eV for monolayer, bilayer, trilayer, and tetralayer phosphorene, in good agreement with the experimental values. The *ab initio* results show that in one through three layer phosphorene, all low energy resonances do indeed arise from inter-subband transitions. In tetralayer and thicker layer phosphorene, however, hybridization between different subbands can lead to band crossings and mixing of higher sub-band transitions (above resonance II); in this case, our simple tight-binding model breaks down.

In summary, our measurements, supported by *ab initio GW-BSE* calculations, demonstrate concretely the extraordinary layer-dependent electronic structure in phosphorene arising from layer-layer interactions. It not only leads to a direct optical bandgap evolving from 1.73 eV in monolayer, 1.15 eV in bilayer, and 0.83 eV in trilayer to ultimately 0.35 eV in the bulk, but also a rich set of sub-bands in few-layer phosphorene that are characterized by strong above-bandgap optical resonances. Together with a high intrinsic electron mobility and the capability to fine-tune the bandgap with mechanical strain and/or electrical field, monolayer and few-layer phosphorene could lead to novel atomically thin electronic and optoelectronic devices over a broad spectral range.

**Acknowledgements:**

L.L., Z.Z., F.Y. and Y.Z. acknowledge support from NSF of China (grant nos. 11425415 and 11421404) and National Basic Research Program of China (973 Program; grant no. 2013CB921902). J.K., C.J. and F.W. acknowledge support from National Science Foundation EFRI program (EFMA-1542741). L.L. and Y.Z. also acknowledge support from Samsung Global Research Outreach (GRO) Program. Part of the sample fabrication was conducted at Fudan Nano-fabrication Lab. G.J.Y and X.H.C. acknowledge support from NSF of China (grant no. 11534010), the 'Strategic Priority Research Program' of the Chinese Academy of Sciences (grant no. XDB04040100) and the National Basic Research Program of China (973 Program; grant no. 2012CB922002). K.W. and T.T. acknowledge support from the Elemental Strategy Initiative conducted by the MEXT, Japan. T.T. also acknowledges support by a Grant-in-Aid for Scientific Research on Innovative Areas, "Nano Informatics" (grant nos. 262480621 and 25106006) from JSPS. D.Y.Q., F.H.J., and S.G.L. thank T. Cao and Z. Li for discussions. The theoretical studies were supported by the Theory Program at the Lawrence Berkeley National Lab through the Office of Basic Energy Sciences, U.S. Department of Energy under Contract No. DE-AC02-05CH11231, which provided for ab initio GW-BSE caclulations, and by the National Science Foundation under Grant No. DMR15-1508412, which provided for DFT calculations and theoretical analyses of interlayer interaction and substrate screening. D.Y.Q. acknowledges support from the NSF Graduate Research Fellowship Grant No. DGE 1106400. This research used resources of the National Energy Research Scientific Computing Center, which is supported by the Office of Science of the U.S. Department of Energy, and the Extreme Science and Engineering Discovery Environment (XSEDE), which is supported by National Science Foundation grant number ACI-1053575."

**Author Contribution:**

F. W., Y. Z. and X. H. C. conceived the project. G. J. Y., X. H. C, L. C. and W. R. grew bulk crystal black phosphorus. L. L. fabricated and characterized the encapsulated samples. J. K. and C. J. designed and built the absorption and PL setup. L. L., J. K. and C. J. obtained and analyzed absorption and PL spectra. C. J. performed the tight binding phenomenological model. D. Y. Q., F. J. and S. G. L performed the ab initio DFT and GW-BSE calculations. Z. S. helped with FTIR measurement and Z. Z. and F. Y. helped with sample fabrication. K. W. and T. T. grew hBN crystal. All authors discussed and wrote the manuscript together.

**Figure captions:**

**Figure 1 | Few-layer phosphorene samples. a,** Puckered honeycomb lattice of monolayer phosphorene. $x$ and $y$ denote the armchair and zigzag crystal orientation, respectively. **b** and **c,** Optical image of few-layer phosphorene samples. Images were recorded with a CCD camera attached to an optical microscope. Number of layers (indicated in the figure) is determined by the optical contrast in the red channel of the CCD image. **d**, Optical contrast profile in the red channel of the CCD images along the line cuts marked in **b** and **c**. Each additional layer increases the contrast by ~ 7%, up to tetralayer.

**Figure 2 | Layer-dependent reflection spectra. a-e,** Reflection spectra of monolayer **a**, bilayer **b**, trilayer **c**, tetralayer **d** and pentalayer **e** phosphorene. Data were obtained at 77 K. Spectra taken under $x$- and $y$-polarized illumination are shown in black and red, respectively. Bandgap of monolayer (1.73 eV), bilayer (1.15 eV), and trilayer (0.83 eV) are determined from the energy of the lowest energy peaks (peak I in black curves). Bandgap transitions of tetralayer and pentalayer are below 0.75 eV and are outside our measurement range. Few-layer phosphorene shows additional peaks (peak II and III in black curves) originating from sub-band transitions. No peak is observed in the reflection spectra for $y$-polarized light. **f,** Reflection spectra of bulk black phosphorus at room temperature.

**Figure 3 | Layer-dependent PL spectra. a-c,** Photoluminescence spectra of monolayer **a**, bilayer **b**, and trilayer **c** phosphorene recorded at 77 K with unpolarized photoexcitation at 2.33 eV. Strong peaks are observed for PL detection in *x*-polarization (black). The peak energy matches well with that in absorption spectra (blue dashed curves), confirming the direct nature of the bandgaps. PL is absent when detected in *y*-polarization (red curves). **d-f,** Intensity of the PL at peak energy as a function of polarization angle $\theta$. Data taken on monolayer, bilayer and trilayer phosphorene are shown red circles in **d, e** and **f**, respectively. All angular dependences show a nearly perfect $cos^2\theta$ pattern (black solid curves), consistent with the fact that optical transitions along *y* direction are forbidden by symmetry.

**Fig. 4 | Evolution of optical resonance energy and band structure. a,** Observed optical resonance energy (solid squares) compared to prediction (solid curves) by the phenomenological one-dimensional tight-binding model. Empty squares are inter-band optical transitions predicted by the tight-binding model that fall outside of our measurement range. The lower and upper limit of our measurement range is marked by horizontal dashed lines. **b,** Calculated GW quasiparticle band structure of monolayer (1L), bilayer (2L), trilayer (3L) and tetralayer (4L) phosphorene. Few-layer phosphorene is encapsulated between a sapphire substrate and a hBN capping layer in the calculation. As the number of phosphorene layers increases, the direct band gap at Γ decreases, and additional high energy sub-bands (marked by I, II and III) emerge.

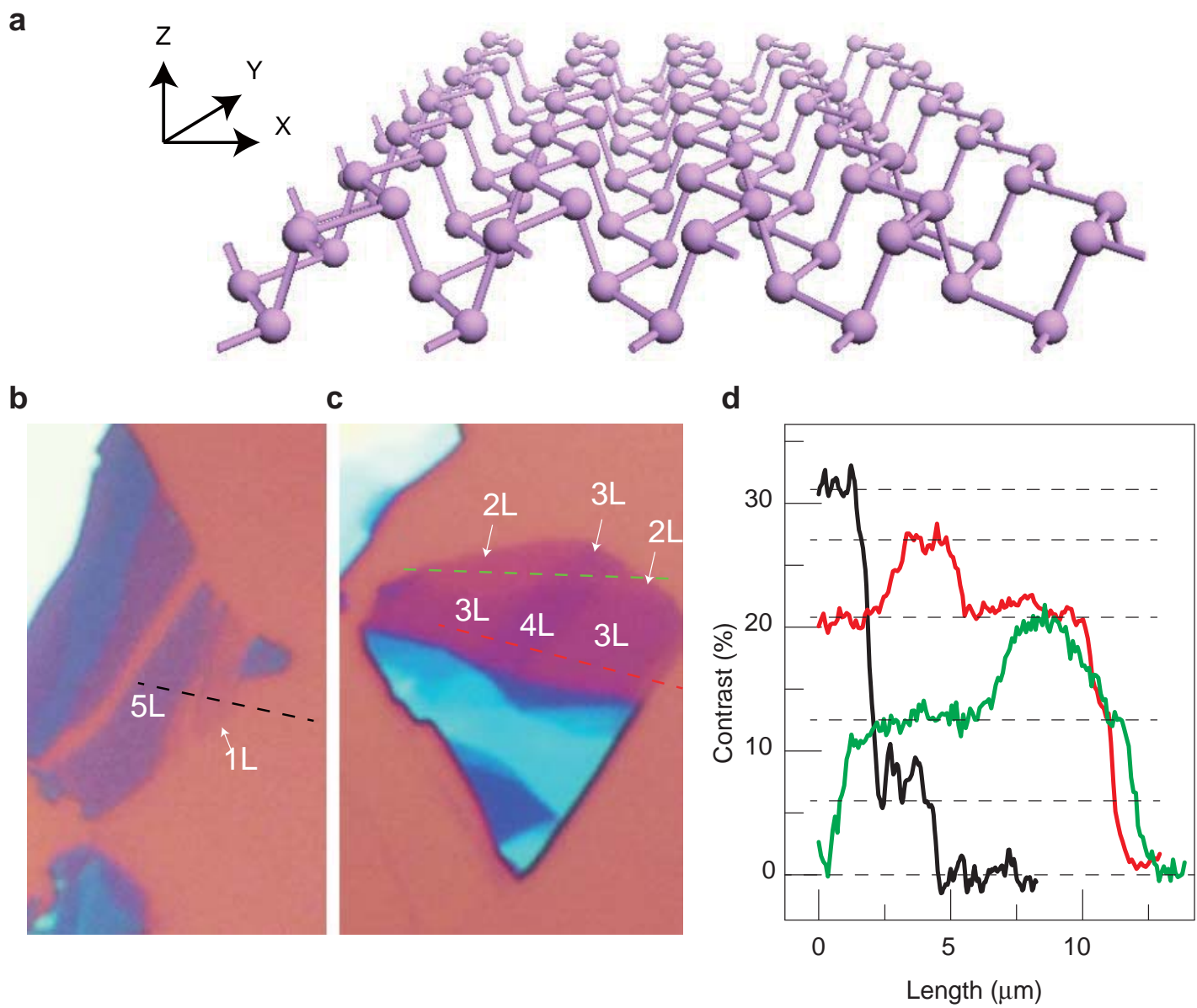

Figure 1

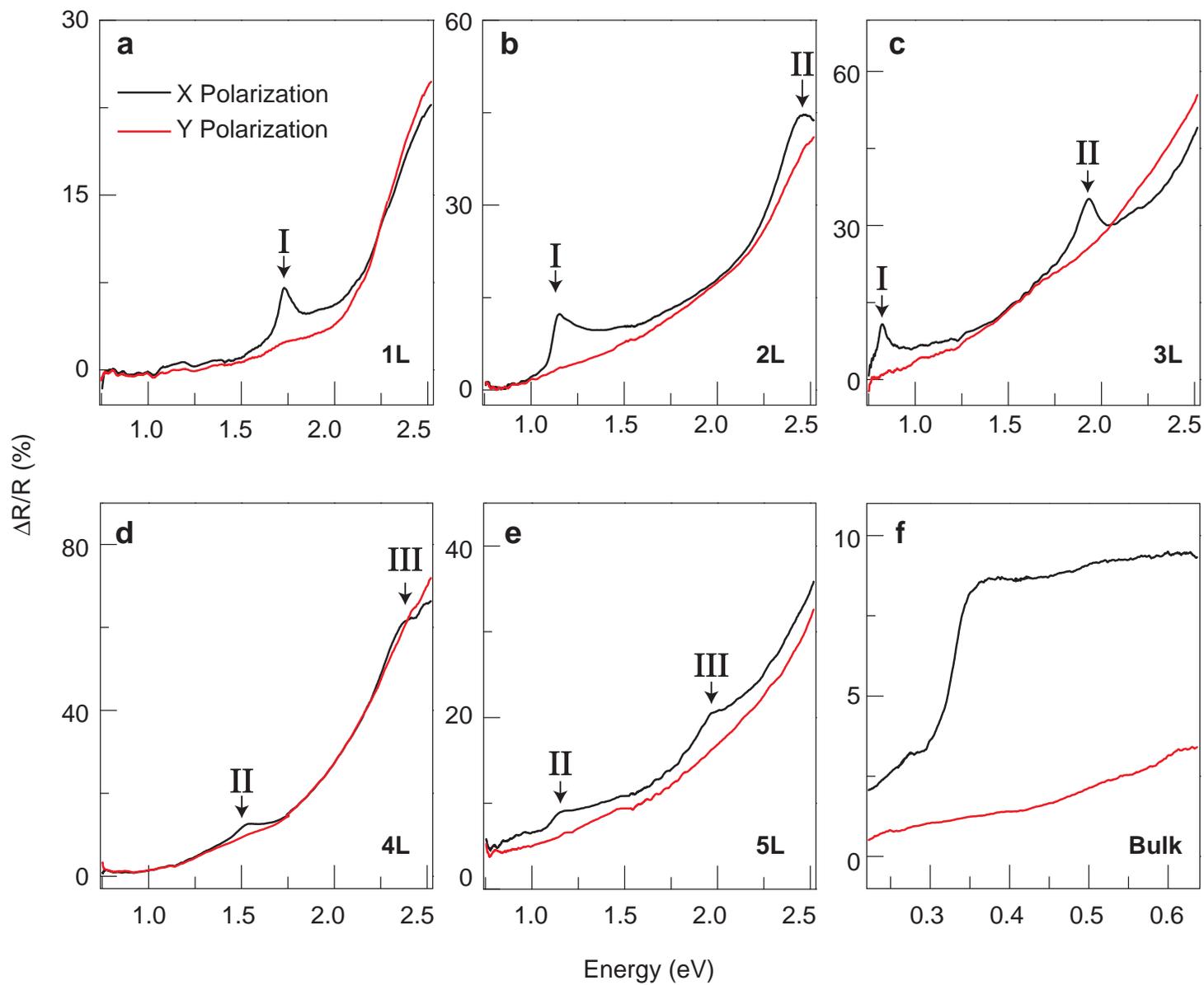

Figure 2

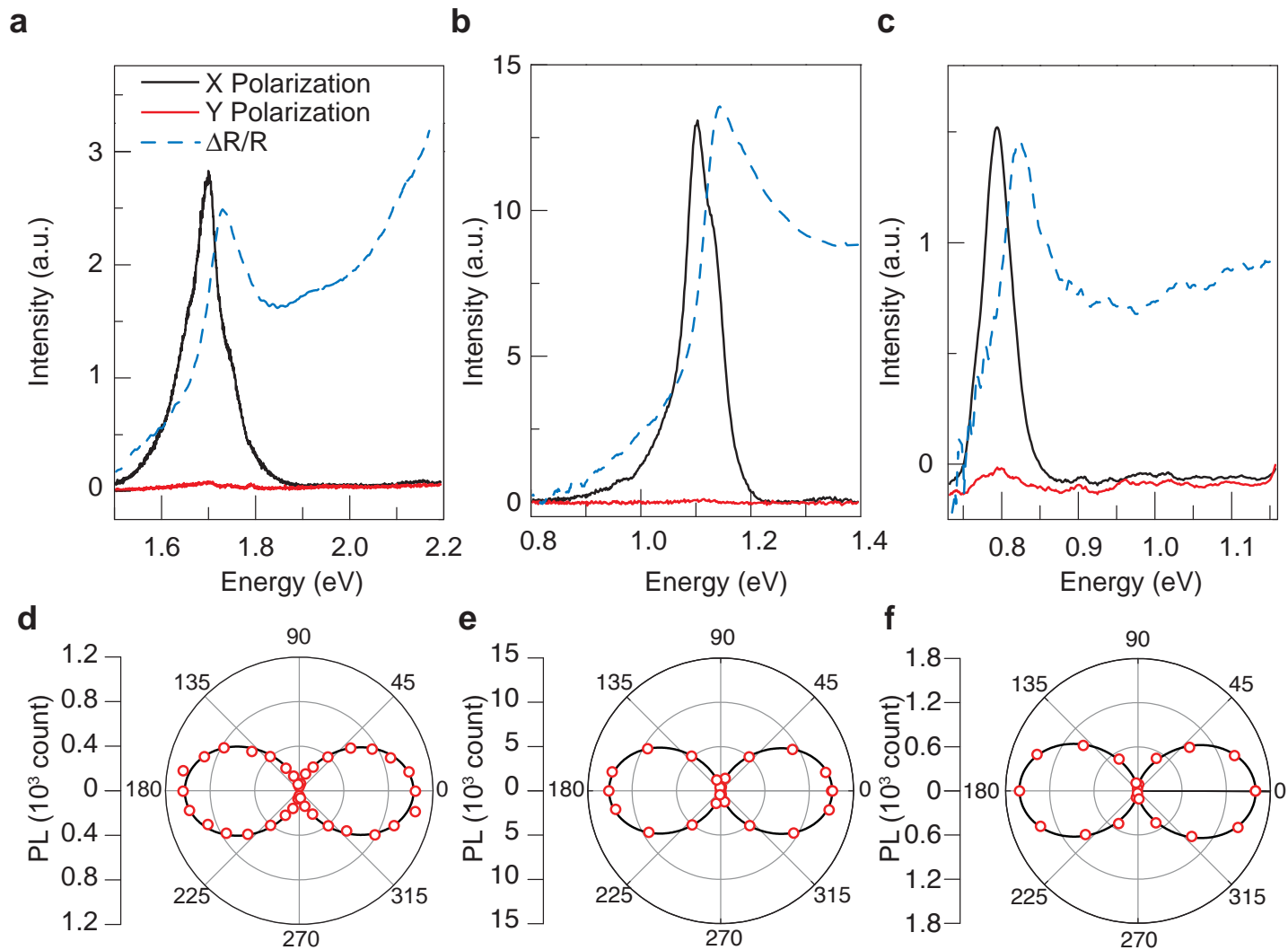

Figure 3

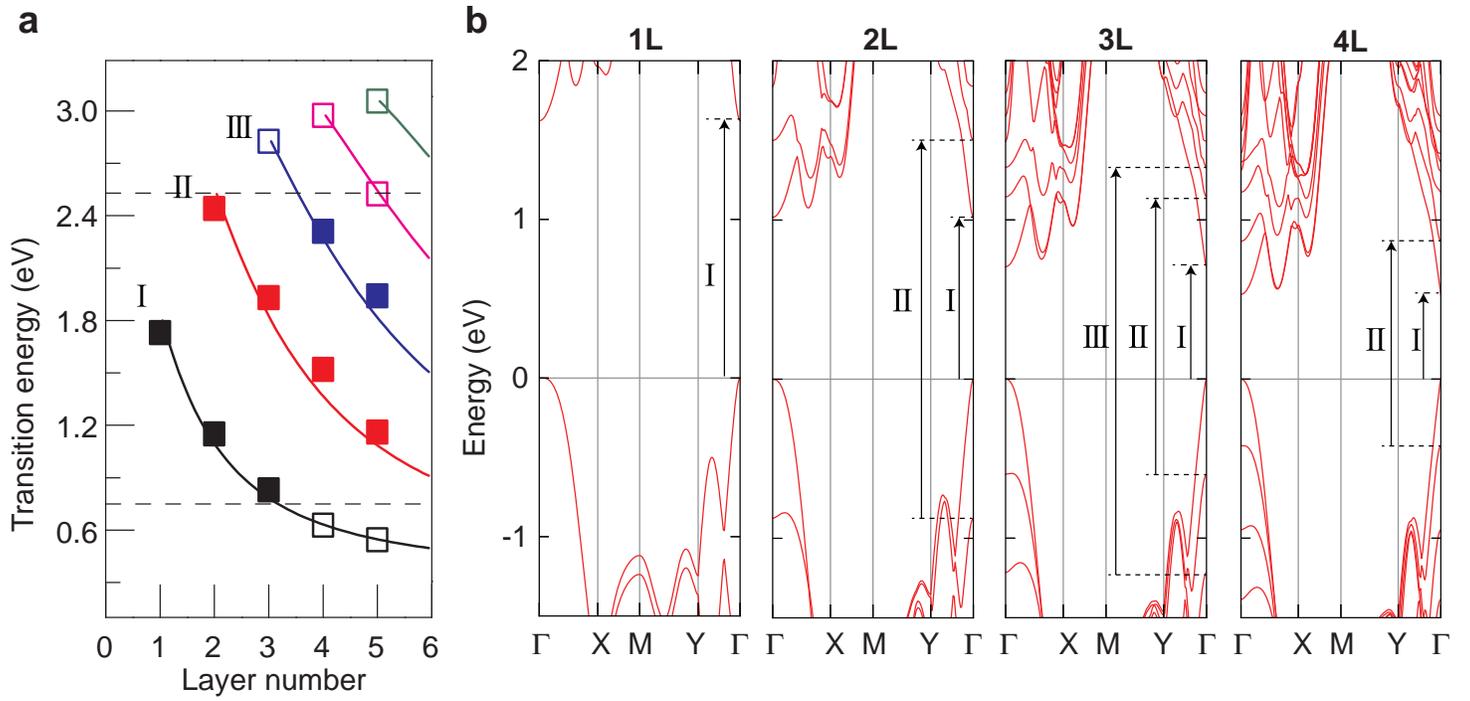

Figure 4